\documentclass[12pt,onecolumn,a4paper, draftcls]{IEEEtran}
\usepackage{graphicx}
\usepackage{epstopdf}

\usepackage[utf8]{inputenc}
\usepackage[T1]{fontenc}
\usepackage[english]{babel}
\usepackage{amssymb}
\usepackage[T1]{fontenc}
\usepackage{hyperref}
\usepackage[english]{babel}
\usepackage{url}
\usepackage{multirow}
\usepackage{float}
\usepackage{cite}
\usepackage{amssymb,amsmath}
\usepackage[]{algorithm}
\usepackage[]{algpseudocode}
\usepackage{multicol,lipsum}
\usepackage{mathtools,esdiff}
\usepackage{booktabs}

\usepackage{soul}  % use \st{works}, where works will be marked with a medium line across the works

\usepackage{bm}
\usepackage{subcaption}
\begin{document}

\title{Machine Learning-based Methods for Joint {Detection-Channel Estimation} in OFDM Systems}

\author{{Wilson de Souza Junior}, {Taufik Abrão}\\
\thanks{This work was partly supported by the National Council for Scientific and Technological Development (CNPq) of Brazil under Grants 310681/2019-7, and in part by the CAPES (Financial code 001), and by State University of Londrina (UEL), PR, Brazil.}
\thanks{Wilson de Souza Junior and T.  Abrão are with the Electrical Engineering Department, State University of Londrina, PR, Brazil.  Rod. Celso Garcia Cid - PR445,  s/n, Campus Universitário, Po.Box 10.011.  CEP: 86057-970. E-mail: \url{wilsoonjr98@gmail.com}; \quad \url{taufik@uel.br}}
}

\maketitle

\begin{abstract} 
In this work, two machine learning (ML)-based structures for  {joint detection-channel} estimation in OFDM systems are proposed and extensively characterized.  Both ML architectures, namely Deep Neural Network (DNN) and Extreme Learning Machine (ELM), are developed {to provide improved data detection performance} and compared with the conventional {matched filter (MF)  detector equipped with the minimum mean square error (MMSE) and least square (LS) channel estimators}.  The bit-error-rate (BER) performance {\it vs} computational complexity trade-off is analyzed, demonstrating the superiority of the proposed DNN-OFDM and ELM-OFDM detectors methodologies.
\end{abstract}

\begin{IEEEkeywords}
Machine learning, neural networks, OFDM, detection, deep learning, DNN, ELM, MMSE, LS, BER.
\end{IEEEkeywords}

\section{Introduction}

%--------------------------
The conventional orthogonal frequency-division multiplexing (OFDM) system is a multi-carrier scheme widely utilized in communication systems due to its capacity to combat frequency-selective fading in wireless channels. Besides, Artificial intelligence (AI) and machine learning (ML) are relevant approaches in the current complex, highly demanded radio access scenarios, combined with the spectrum scarceness. The ML resources and techniques can be applied to improve the performance-complexity trade-off of OFDM systems, specifically on the receiver side. In this work, two AI-based methods, specifically a DNN-based and an ELM-based {jointly symbol detection and pilot-assisted channel estimation are investigated}; both techniques are compared with the conventional linear estimation methods, such as least square (LS) and minimum mean square error (MMSE) \cite{cho,james2011channel}. 

ML techniques have been widely used in different telecommunication applications as a satisfactory predictor in OFDM system \cite{1,ELM1,ELM2}, as a near-optimal signal detection in OFDM with index modulation (OFDM-IM) \cite{2}, and as a channel estimator for massive MIMO \cite{massMIMO}.  In \cite{1}, the authors discuss the deep learning (DL) applicability for channel estimation and signal detection in OFDM systems.  The DL-based prediction technique is explored to implicitly estimate the channel state information (CSI) and then detect the transmitted symbols using the estimated CSI. For that, the deep learning model is first trained offline using the data generated by simulation based on channel statistics and then used for recovering the online transmitted data directly. 

Deep neural network (DNN) is a type of artificial neural network presenting a large number of hidden layers and hyper-parameters into its composition, {\it i.e.}, DNN implies high computational operations in contrast to Extreme Learning Machine (ELM) that has simply one hidden layer \cite{neuro}. In intricate telecommunication scenarios, specifically in the 5G and beyond systems, authors in \cite{ELM1} propose a different architecture for the OFDM receiver aided by ELM technique. Besides, a multi-ELM, {\it i.e.,} a parallel multiple split complex ELM structure is proposed in \cite{ELM2}. 
In \cite{2}, a DL-based detector structure for OFDM with index modulation (OFDM-IM) is proposed, termed DeepIM. The authors deploy a deep neural network with fully connected layers to recover data. Aiming to enhance the DeepIM performance, the received signal and channel vectors are pre-processed based on the domain knowledge before entering the network. Data sets available by simulations are deployed to train offline the DeepIM  aiming at optimizing the bit error rate (BER) performance. After that, the trained model is deployed for the online signal detection.

\vspace{2mm}
\noindent\textit{\textbf{Contributions}}.  
We propose and analyze the deployment of promising ML tools applied through jointly detection and channel estimation in OFDM systems. {\bf i}) First, we have adopted and analyzed two ML-based topologies for OFDM data detection and channel estimation: a DNN-based and ELM-based OFDM joint detector and channel estimator. {\bf ii}) We have deployed and characterized {existing models by applying them to more complex scenarios, including multi-user systems, realistic} path-loss, and short-term fading wireless channel configurations.  {\bf iii}) Extensive numerical results characterizing the performance-complexity trade-off for both ML-based OFDM detectors, demonstrating that such an approach is quite competitive. 

\vspace{1mm}
\noindent\textbf{\textit{Notations}}.
 {Italic lowercase or capital letters are scalars}, boldface capital letters denote the frequency-domain vectors meanwhile boldface lowercase letters are vectors in the time domain. Operators {$\mathbb{E}[\cdot]$,} $(\cdot)^T$, $(\cdot)^H$ and $(\cdot)^\dagger$ {denote the statistical expectation,} a vector {or} matrix transpose, Hermitian and {Moore-Penrose} pseudo-inverse, respectively; $|\mathcal{A}|$ holds for the cardinality of the set $\mathcal{A}$, $\odot$  denotes the element-wise multiplication, {$\oslash$  denotes the element-wise division,} and $\circledast$ {represents} the convolution operation.
 
%------------------------------------
\section{System Model}\label{sec:sys_model}
%------------------------------------
 \vspace{-.2cm}
 Assuming an OFDM system with {a set $\mathcal{U}=\{1,\dots,U\}$ users}, {a number of} $N_{c}$ sub-carriers and the duration of the cyclic prefix $T_g$, a transmitted signal {in frequency domain} can be defined as $\textbf{X} = [X_1, \dots , X_{N_{c}}]^T$, leading to a received signal $\textbf{Y}= [Y_1,\dots, Y_{N_{c}}]^T$, with multi-path channel $\textbf{H} = [H_1, \dots , H_{N_{c}}]^T$, and zero-mean Gaussian noise samples  {$\textbf{Z} = [{Z}_1, \dots , Z_{N_{c}}]^T$} described by complex random variables {$Z_i \sim \mathcal{C}\mathcal{N}(0,\sigma^2)$}, where $\sigma^2$ is the noise power {in each OFDM sub-channel}. The received signal in the {\it frequency} and {\it time domain} can be written, respectively: 
\begin{equation}
\textbf{Y} = \textbf{X} \odot  \textbf{H} + {\textbf{Z}}, \qquad \text{{and}} \qquad \textbf{y} = \textbf{x} \circledast   \textbf{h} + {\textbf{z}}
\end{equation}
where \textbf{y}, \textbf{x}, \textbf{h} and {\textbf{z}} means IDFT of {\textbf{Y}}, {\textbf{X}}, {\textbf{H}} and {\textbf{Z}} respectively. In the considered path-loss model the received signal power decays according to $d_k^\eta$, where $d_k$ is the distance between BS and the $k$th user, while  $\eta$ represents the path-loss exponent. {Hence, the transmitted power per sub-carrier and the average receiver power per subcarrier ($P$) are related by: $P= d_u^{-\eta}\cdot \frac{P_{\textsc{t}}}{N_{c}}$, where {$P_{\textsc{t}}$ is the total power available at transmitter side}.
Besides, since more than one user sharing the same sub-channel is admitted, resulting in an OFDM system operating under inter-user interference (IuI), in Section \ref{sec:scsel} we proceed with sub-carrier selection to know what sub-carriers sub-set results in smaller IuI, aiming to maximize
the SINR  in the $k$th sub-carrier.
}
 
 \vspace{2mm}
%--------------------------
\noindent{\bf LS OFDM Channel Estimation}. 
%--------------------------
As aforementioned, firstly we assume that an OFDM pilot symbol is transmitted (channel estimation mode) and then OFDM data symbols can be transmitted (data mode) inside the channel coherence time $(\Delta t)_\textsc{c}$ interval; this composition form an OFDM frame. Inside an OFDM frame, the channel state information (CSI) is unchangeable but it changes from one frame to another. One common technique for OFDM channel estimation is {the {\it least-squares} (LS) method} \cite{cho}. This technique is the simplest way to estimate the state of the channel; as a result, it is possible to estimate the OFDM symbols inside the same $(\Delta t)_\textsc{c}$ time interval. Once {$\textbf{X}_p= [X_1,\dots, X_{N_{c}^{\rm pilot}}]^T$ and $\textbf{Y}_p= [Y_1,\dots, Y_{N_{c}^{\rm pilot}}]^T$} are the transmitted and received OFDM pilot
{vectors}, respectively, {with $N^{\rm pilot}_{c}$ the number of sub-carriers reserved to the pilots in the OFDM frame,} then the LS {\it channel estimation} {in the pilot OFDM sub-channels, and the {\it data detection} based on LS channel estimator are obtained, respectively, by}:
\begin{equation}
\tilde{\textbf{H}}^{\rm{LS}} = \textbf{Y}_p\oslash \textbf{X}_p, \qquad \text{and}\qquad \tilde{\textbf{X}}_d = \textbf{Y}_d \oslash \tilde{\textbf{H}}^{\rm{LS}}, 
\end{equation}
where $N^{\rm data}_{c}$ is the number of sub-carriers destined to data symbol in the OFDM frame; $\tilde{\textbf{X}}_d$ and $\textbf{Y}_d$ are the recovery data and received data {vectors}, respectively.

 \vspace{2mm}
%--------------------------
\noindent{\bf MMSE OFDM Channel Estimation.}
%--------------------------
The MMSE channel estimator is considered a better linear solution than the aforementioned LS channel estimation due to the weight (regularization) channel matrix inversion, which is optimized in the same way as the LS solution according to the minimum mean square error problem. However, the development of the MMSE solution requires the knowledge of the {signal-to-noise ratio (SNR)}, being the channel estimate obtained from the LS solution as \cite{cho}:
\begin{equation}
    \tilde{\textbf{H}}^{\rm{MMSE}} = \mathbf{R}_{\mathbf{H} \tilde{\textbf{H}}^{\rm{LS}} } \begin{bmatrix}  \mathbf{R}_{\mathbf{HH}} + \mathbf{I} \frac{1}{\Bar{\gamma}} \end{bmatrix}^{-1}   \tilde{\textbf{H}}^{\rm{LS}}, 
\end{equation}
where $\mathbf{R_{AB}}$ denotes the cross-correlation matrix between matrices \textbf{A} and \textbf{B}, $i.e$., $\mathbf{R_{AB}} = \mathbb{E}[\textbf{AB}^H]$; the pre-processing SNR at the receiver side is defined as $\Bar{\gamma} \triangleq \frac{P}{\sigma^2}$, with {$P$ the average power per sub-channel at receiver side}.

 \vspace{2mm}
%====================================
\section{ML-based OFDM Detection Schemes}\label{sec:DL-OFDM_detectors}
\begin{figure}[!htbp]
\centering
\begin{minipage}{.48\textwidth}
  \centering
  \includegraphics[trim={1mm 1mm 1mm 1mm},clip,width=1\linewidth]{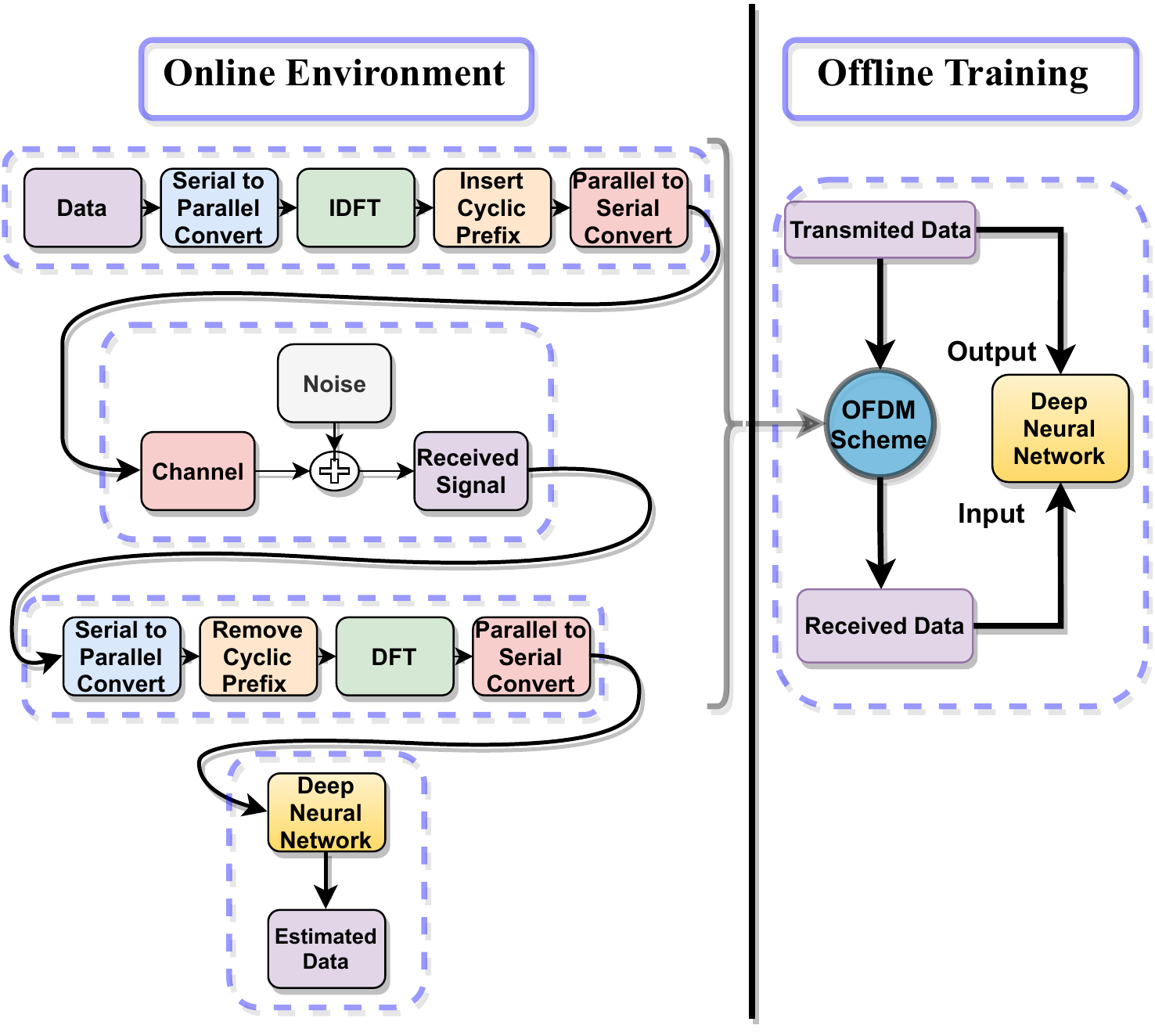}
  \captionof{figure}{\footnotesize OFDM \& DNN training.}
  \label{fig:sys}
   \vspace{-4mm}
\end{minipage}
\quad
\begin{minipage}{.48\textwidth}
  \centering
  \includegraphics[width=1\linewidth]{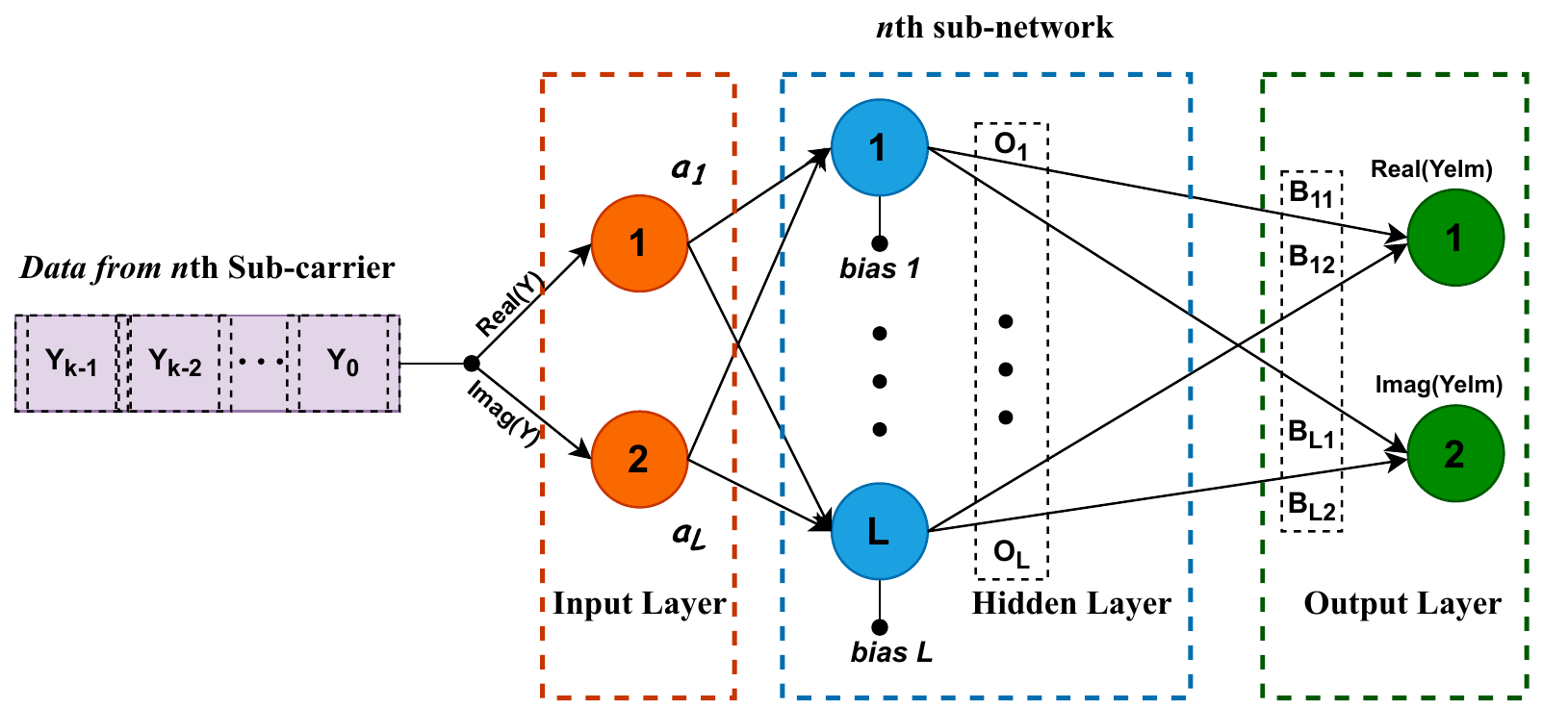}
  \captionof{figure}{\footnotesize DNN architecture.}
  \label{fig:DNN}
   \vspace{-4mm}
\end{minipage}%
\quad
\begin{minipage}{.48\textwidth}
  \centering
  \includegraphics[trim={1mm 1mm 1mm 1mm},clip,width=1\linewidth]{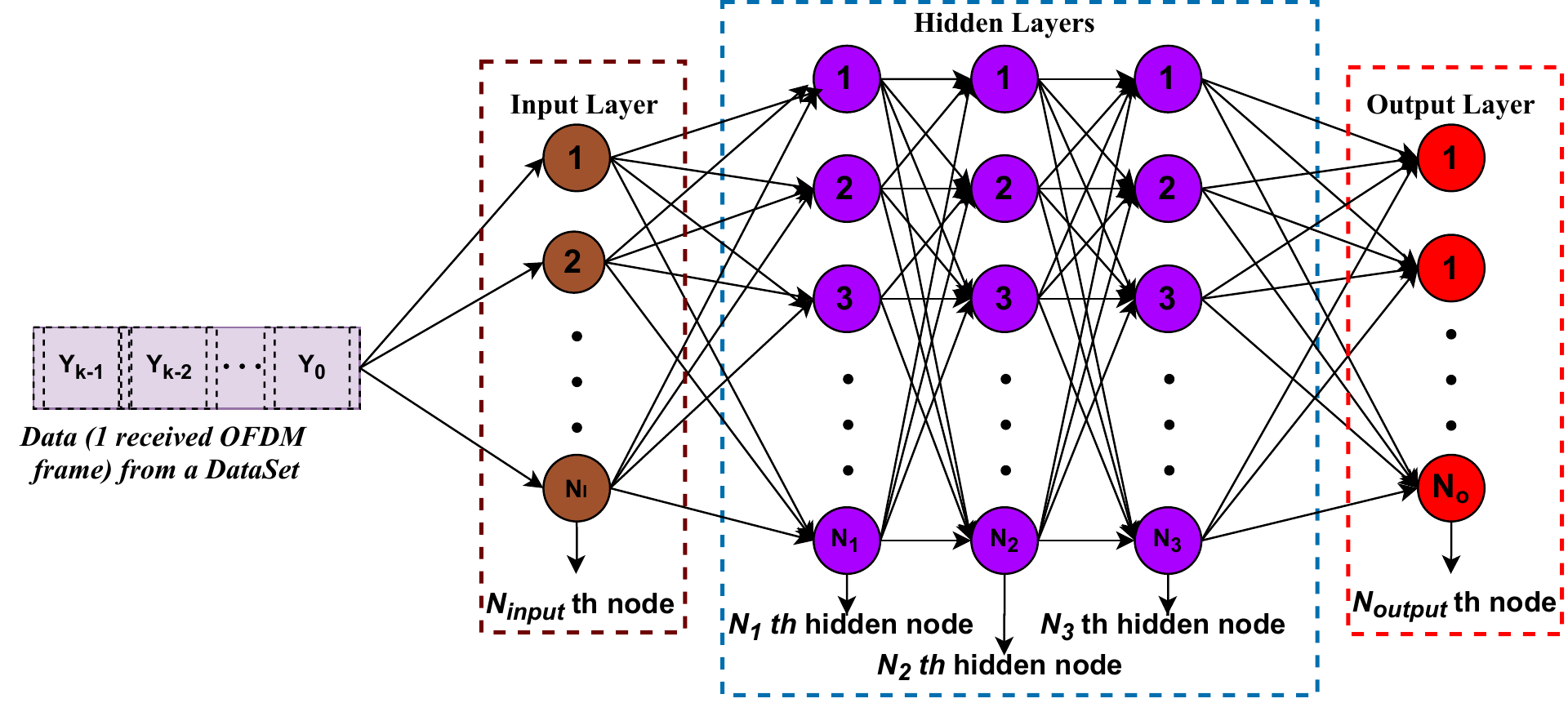}
  \captionof{figure}{\footnotesize ELM architecture.}
  \label{fig:ELM}
\end{minipage}
\begin{minipage}{.48\textwidth}
  \centering
  \includegraphics[width=1\linewidth]{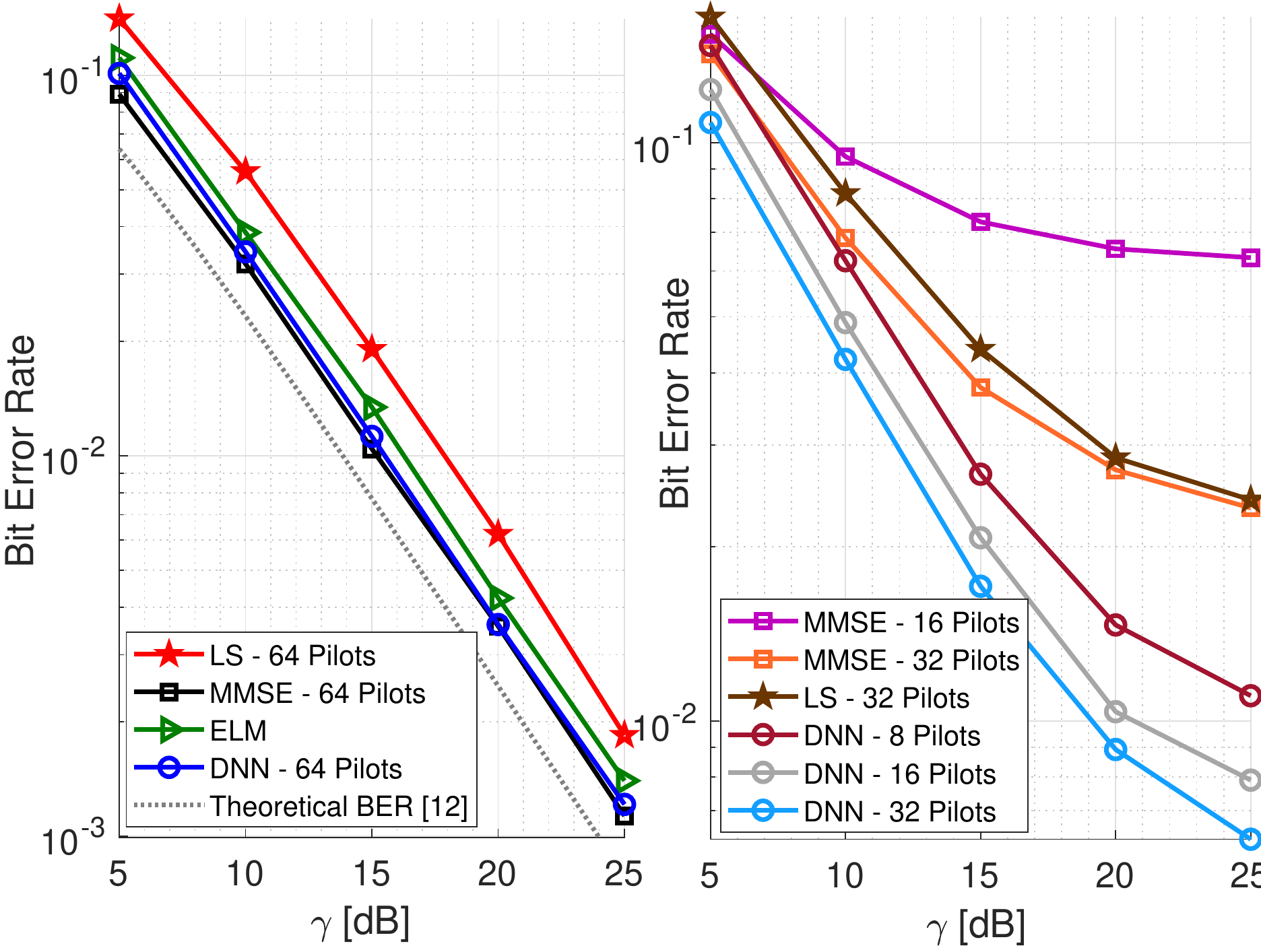}
  \captionof{figure}{\footnotesize BER for DNN, LS and MMSE CE under 4-QAM: (a) 64 pilots  and (b) Reducing \#pilots}
  \label{i:2}
\end{minipage}%
\hspace{.2cm}
 \vspace{-4mm}
\end{figure}

%====================================
 \vspace{.2mm}
In the context of machine learning, the training occurs by generating random data communicating across a channel that arrives at the receiver, and then that data is part of a data set containing labels and features. This paper presents an analysis and comparison of two different ML-based detectors that are promising for realistic OFDM system scenarios. {The deployed OFDM system model is depicted in Fig. \ref{fig:sys}; the} DNN and ELM architectures are described in the following.

\vspace{.1mm}
%--------------------------
\noindent{\bf DNN-based Detection}. 
%------------------------------------
The DNN is an architecture composed by $2\cdot N_c$ input nodes being a real and imaginary part of OFDM frame, where {$N_c = N^{\rm pilot}_{c} + N^{\rm data}_{c}$}. This model has exclusively 3 hidden layers, Fig. \ref{fig:DNN}, in which each layer is composed of 500, 250, and 120 neurons, respectively, in the same way as adopted in \cite{1}.

The proposed DNN-based OFDM detector is exclusively inspired in offline training strategy, {\it i.e.}, for the training stage, the DNN inputs (features) are composed by the real and imaginary part of OFDM symbols that arrives at the receiver, while the outputs (labels) are estimates for the transmitted bits.  
 The DNN outputs are obtained from a non-linear function of input nodes: 
\begin{equation}\label{eq:Xdnn}
\mathbf{\hat{X}}_{\textsc{dnn}} = f(\textbf{Y}, \mathbf{\theta }) = f^{L-1}(f^{L-2}(...f^{1}(\mathit{\mathbf{Y}})))
\end{equation}
where $\mathbf{\theta}$ is the set of bias and weights and $L$ means the number of layers. 
The bias and weight coefficients are optimized in the training stage.  
The DNN model has the goal of minimizing the  mean squared error (MSE) loss function, defined by:
\begin{equation}
    \mathcal{F}_{\rm loss} = \frac{1}{L} \sum_{k=1}^{L}\left[\mathit{\mathbf{\hat{X}}}(k) - \mathit{\mathbf{X}}(k)\right]^2
\end{equation}
where $\mathit{\mathbf{\hat{X}}}(k)$ denotes the predictions, $\mathit{\mathbf{X}}(k)$ the data symbol, and $L$ is the number of data samples in the estimation data set.

\vspace{1mm}
\noindent{\it DNN Training}. The training step is responsible for the DNN to learn the channel characteristics; hence the data set must be known and sufficiently large,  beyond it should be transmitted in a fraction of the channel coherence time $(\Delta t)_\textsc{c}$ interval to allow the system attains suitable accuracy in the channel estimate process. Once trained, the network may be utilized to decode data to any online transmission scheme ({\it test stage}), also assuming the same parameters utilized previously at the training stage. 

%--------------------------
\noindent{\bf Extreme Learning Machine (ELM) based Detection}. 
%-----------------------------
The ELM network has other important feature that differs from DNN, its architecture is subdivided into $N_{c}$ sub-networks, {\it i.e.}, in the OFDM context, a sub-network is deployed to treat the signal of each sub-carrier. Fig. \ref{fig:ELM} depicts an ELM topology for each OFDM sub-carrier. The parameters of hidden nodes of ELM must be randomly generated, and afterward, they should be fixed to determine the output layer weights according to \cite{ELM1,ELM2}. ELM architecture has 2 input nodes, the real and imaginary parts of the $n$th sub-channel. In OFDM systems, the ELM architecture assumes that the data and pilots are time-division multiplexed, {\it i.e.}, inside a channel coherence time $(\Delta t)_\textsc{c}$ interval, there are $I$ transmitted pilots and $K$ transmitted data symbols. This is a primordial feature once the channel can be assumed invariant into the $(\Delta t)_\textsc{c}$ interval. Still, the noise and possibly co-channel interference at the receiver side are variant in time. The data set provided by $I$ pilots is helpful for the network to learn about statistics from noise plus interference.

The hidden layer from ELM has an activation function applied to the data from the input layer, Fig. \ref{fig:ELM}, where the output of $\ell$th hidden node is given by:
\begin{equation}
    {\bf o}_{i,L} = g(\textbf{{a}}_i^{T} \cdot \textbf{{Y}}_i + b_L)
\end{equation}
where $L$ means the number of hidden neurons, $\textbf{\textit{a}}_i$ is a column vector of weights with dimension 2 $\times $ 1, while  $b_L$ means the bias from $\ell$th hidden node and $\textbf{{Y}}_i \in \mathbb{R}^{2\times1}$ is the transmitted data. The ${j}$th hidden layer matrix ${\bf O}\in \mathbb{R}^{I\times L}$ is given by:
\begin{equation}
{\bf O}_{j} 
= \begin{bmatrix}
g(\textbf{{a}}_1^{T} \cdot \textbf{{Y}}_1 + b_1) \cdots  g(\textbf{{a}}_L^{T} \cdot \textbf{{Y}}_1 + b_L) \\ \vdots \, \, \, \, \, \, \, \, \, \, \, \, \, \, \, \cdots \, \, \, \, \, \, \, \, \, \, \, \, \, \, \,  \vdots\\
g(\textbf{{a}}_1^{T} \cdot \textbf{{Y}}_I + b_1) \cdots  g(\textbf{{a}}_L^{T} \cdot \textbf{{Y}}_I + b_L)\\
\end{bmatrix}
\end{equation}

Different from DNN, the ELM topology can be trained in a non-iterative mode in order to minimize a training error function:
\begin{equation}\label{eq:Bhat}
\hat{\bf B} = \underset{{\bf B}\in \mathbb{R}^{L\times 2}}{\min} 
\begin{Vmatrix}
\mathbf{O} \mathbf{B} - \mathbf{X}_{\rm pilot}
\end{Vmatrix}  \,\, = \,\,  \mathbf{O}^\dagger \mathbf{X}_{\rm pilot}
\end{equation} 
where $\mathbf{\hat{B}}$ is a $L \times 2$ dimensional matrix denominated {\it output weight matrix}. Once the training stage is completed for each sub-network (sub-carrier), then ELM can operate, {\it i.e.} the ELM-OFDM detector can estimate the data by:
\begin{equation}
\mathbf{\hat{X}}_{\textsc{elm}} = \mathbf{O}{\hat{\mathbf{B}}}
\label{e:1}
\end{equation}

Eqs. \eqref{eq:Xdnn} and \eqref{e:1} explicitly provide estimates of the transmitted signal (OFDM data detection). Although the OFDM channel estimation intrinsically occurs inside the ML detection step, explicitly, we do not proceed with the channel estimation step. In this sense, the proposed ML-based techniques are also suitable for non-coherent signal detection since such schemes explicitly neither require channel state information knowledge at the transmitter nor the receiver.

%==========================
\section{Sub-Carrier Selection for maximizing SINR}\label{sec:scsel}
%==========================
We describe a method for sub-carrier selection and allocation aiming at maximizing the signal-to-interference plus noise ratio (SINR) across the users. Hence, in this work, we fix a pre-defined value for the maximum number of sub-carriers per user, {$N_{\rm cpu}$}, and the number of sub-carriers that can support two OFDM users simultaneously subject to equal interference level in those subcarriers.  Notice that in the adopted system model, we have admitted more than one user sharing the same sub-channel, resulting in an OFDM system operating under inter-user interference. Also, to facilitate the analysis, but without loss of generality, we have defined and selected an equal number of sub-carriers per user that generate the lesser interference over other users, $N_{\rm cpu}^{\rm eq}$. Our principal goal is to allocate as best as possible the sub-carriers among the users in such a way that maximizes the SINR (optimization metric) in the $k$th sub-carrier subject to interference {in a given OFDM frame}, which can be written as:
\begin{align}\label{eq:SINR}
 {\rm SINR}_u(k)  = \,\, & \frac{P_{u}(k) |H_u(k)|^2}{P_{j}(k) |H_j(k)|^2 + \sigma^2}, \qquad j\neq u; \qquad \forall  k= 1,\dots, N^{\rm data}_{c},\quad \forall  u \in \mathcal{U}; 
\end{align}
\vspace{-9mm}

\begin{align*}
\text{with} \quad P_{u}(k)=\frac{P_{\textsc{t}}}{N^{\rm data}_{c}}, \,\,\, \text{and} \,\,\, P_j(k) = \left\{\begin{matrix}
\frac{{P_{\textsc{t}}}}{N^{\rm data}_{c}}, & \text{if}\,\,\,  j \in \mathcal{J}_u\\
0 & \rm otherwise
\end{matrix}\right.
\end{align*}
where $\mathcal{J}_u$ represents the sub-set of users interfering in the $k$th subcarrier of user $u$. Notice that  $P_{u}(k)=\frac{{P_{\textsc{t}}}}{N^{\rm data}_{c}}$ indicates equal power allocation (EPA) policy across the users, where $P_{j}{(k)}= \frac{{P_{\textsc{t}}}}{N^{\rm data}_{c}}$ means the EPA policy also for the $j$th interfering user in the $k$th sub-carrier; $H_u(k), H_j(k)$ mean the channel response for the $u$th and $j$th user in the $k$th sub-carrier, respectively. A pseudo-code for subcarriers selection aiming at maximizing the SINR, eq. \eqref{eq:SINR}, is depicted in Algorithm 1. {$\mathcal{U}\setminus\mathcal{J}_u$ is the set difference; {\it i.e.}, it is the set of all those elements that are in $\mathcal{U}$ but not in $\mathcal{J}_u$}.

\begin{algorithm}[!htbp]
\caption{{Sub-carrier Selection for SINR maximization}}
\begin{algorithmic}[1]\label{alg:Scsel}
\For {$u=1,2,\dots,{|\mathcal{U}\setminus\mathcal{J}_u|}$}
\For {$k$= $N_{\rm cpu}(u-1) + 1$ $:$$N_{\rm cpu}(u-1)+N_{\rm cpu}$} 
\State Evaluate the $k$th SINR for $u$th user as in (\ref{eq:SINR});
\EndFor;
\State  Sort all SINR's for $u$th user according to the descending order; 
\State  Select the first $N_{\rm cpu}^{\rm eq}$ SINR's;
\EndFor
\end{algorithmic}
\end{algorithm}

%............................
\section{Simulation Results}\label{sec:simul}
\begin{figure}[!htbp]
\centering
\begin{minipage}{.48\textwidth}
  \centering
  \includegraphics[trim={1mm 1mm 1mm 1mm},clip,width=1\linewidth]{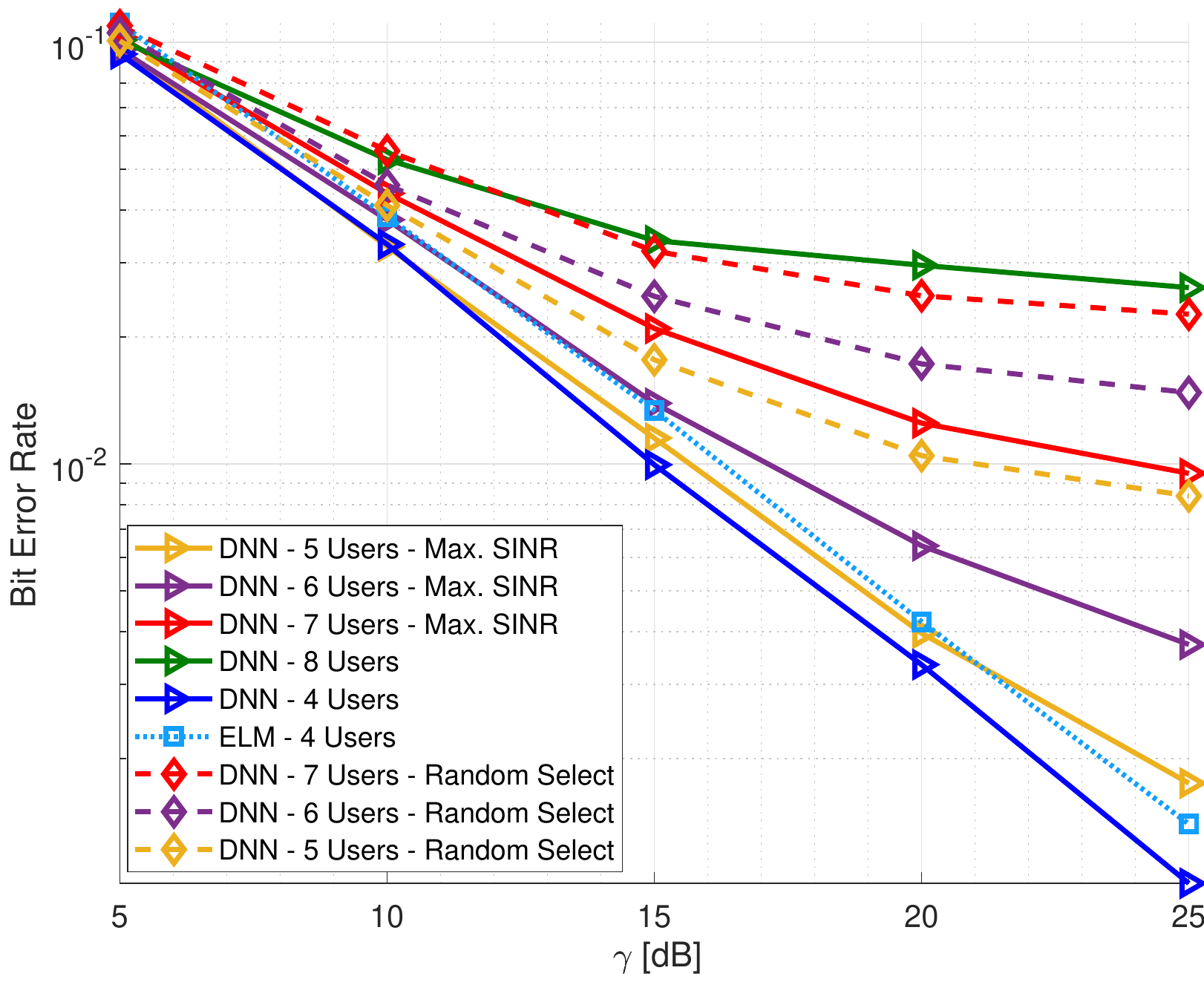}
  \captionof{figure}{\footnotesize BER for DNN with the number of users varying from $U=4$ to $8$ and ${\mathcal{J}_u} = \{5,6,7,8\}$. Modulation: 4-QAM.}
  \label{fig:difuser}
\end{minipage}
\hspace{.2cm}
\begin{minipage}{.48\textwidth}
  \centering
  \includegraphics[trim={1mm 1mm 1mm 1mm},clip,width=1\linewidth]{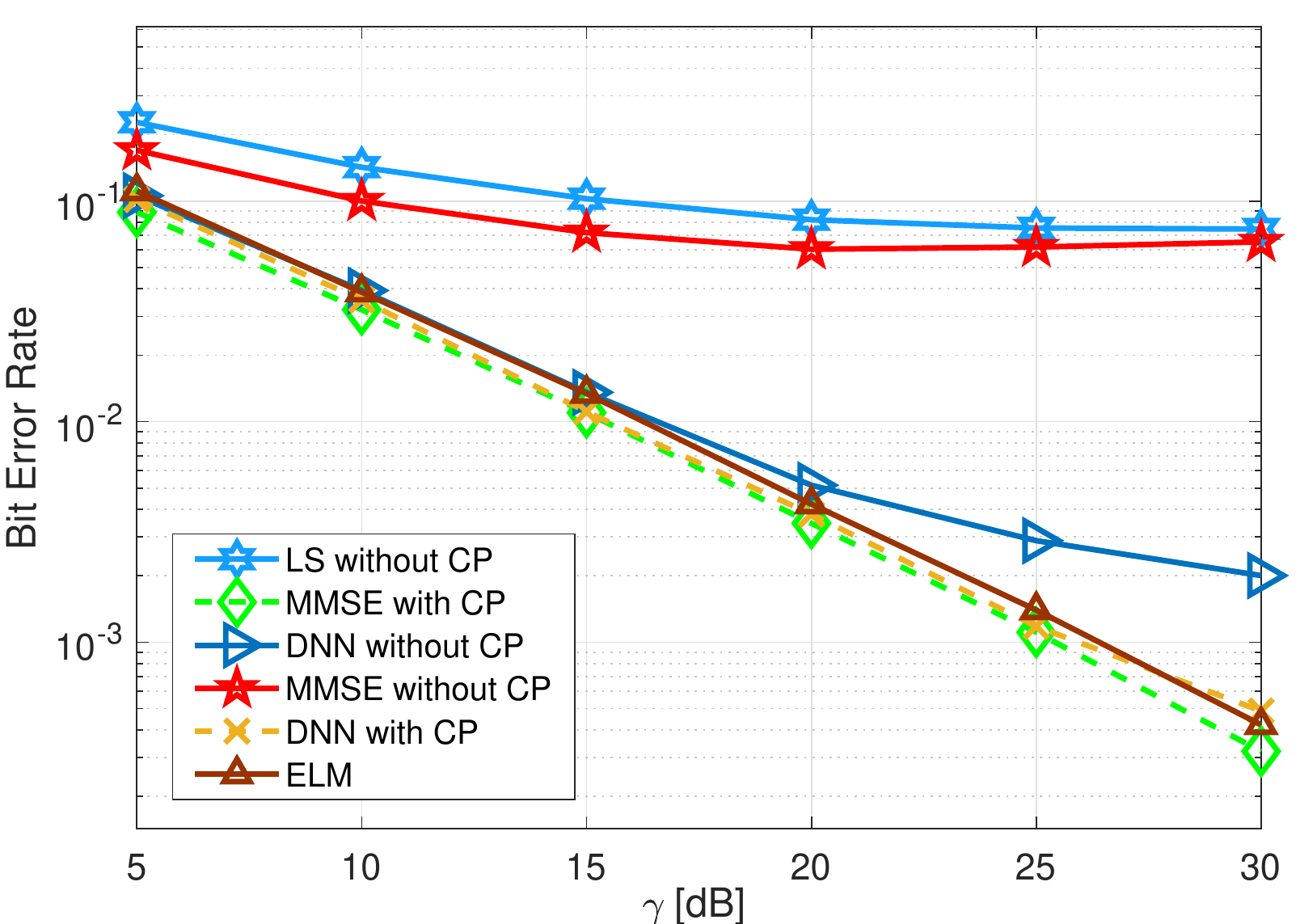}
  \captionof{figure}{\footnotesize Cyclic prefix influence on the BER performance for DNN, MMSE and LS channel estimators for CP = 25\%.}
  \label{i:3}
\end{minipage}
\hspace{.2cm}
\begin{minipage}{.48\textwidth}
  \centering
  \includegraphics[trim={1mm 1mm 1mm .5mm},clip,width=1\linewidth]{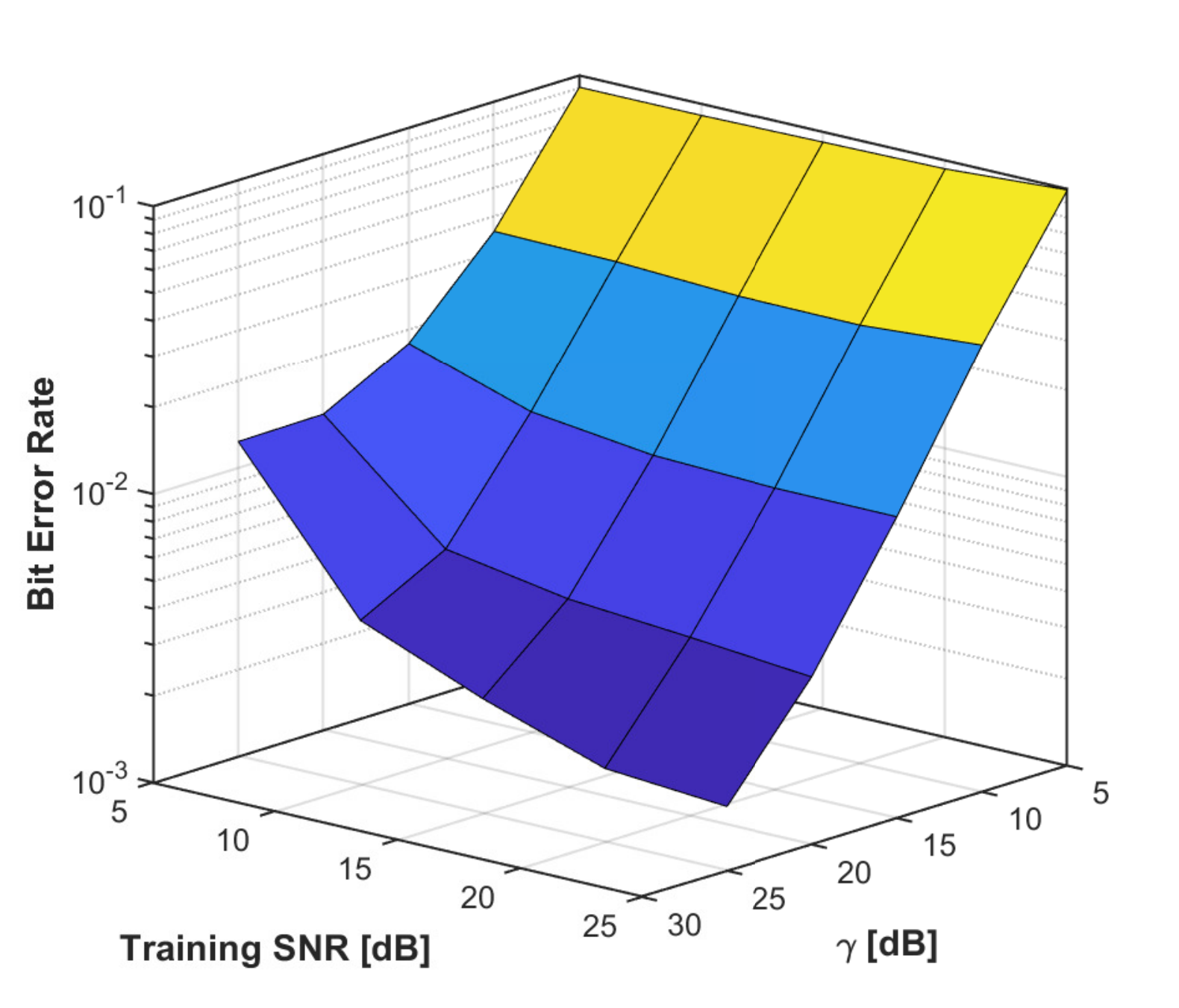}
  \captionof{figure}{\footnotesize BER performance for different range of $\textsc{snr}_{\rm train}$ and CP = 25\%.}
  \label{i:4}
\end{minipage}
\begin{minipage}{.48\textwidth}
 \centering
\includegraphics[trim={1mm 1mm 1mm 1mm},clip,width=1\linewidth]{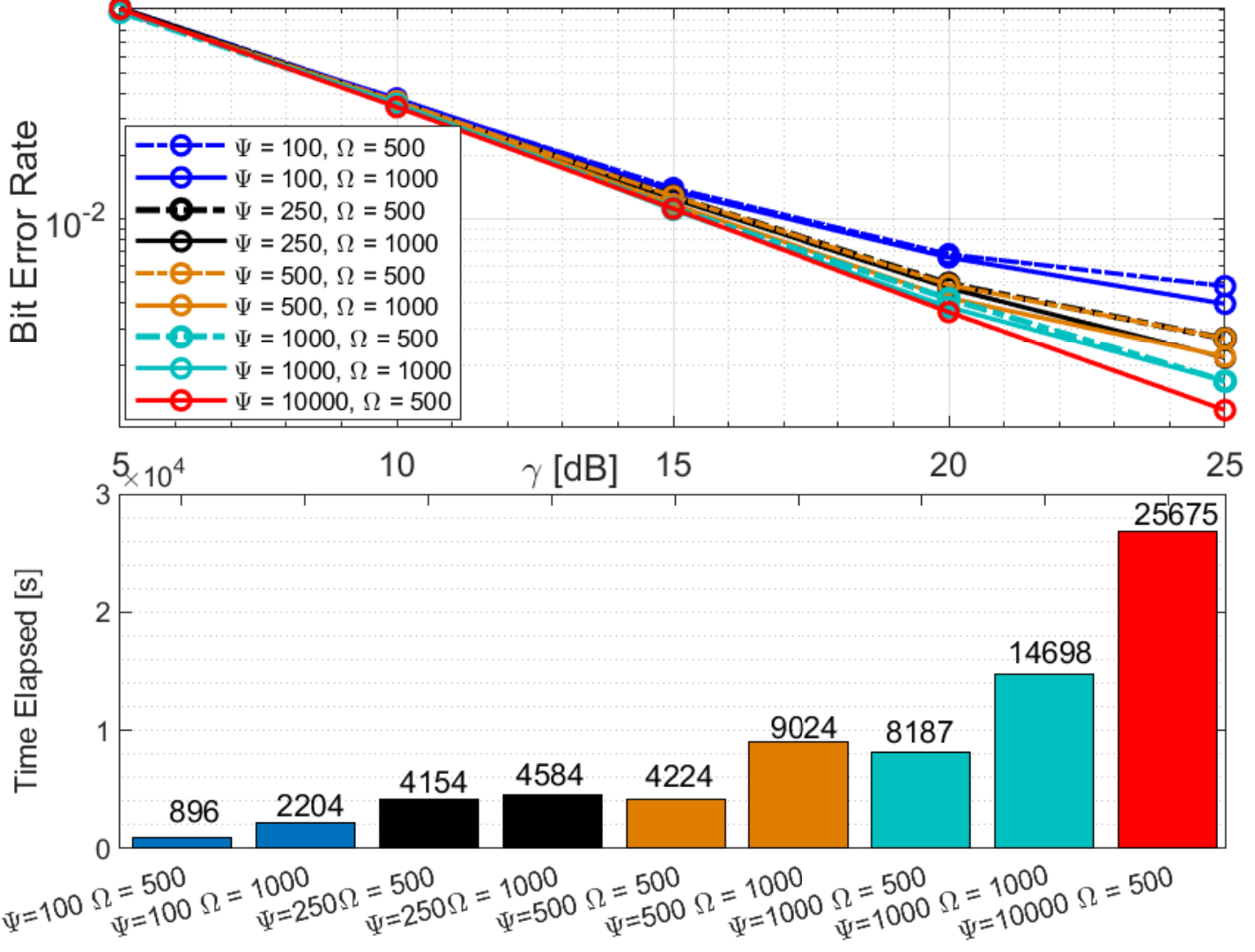}
\captionof{figure}{\footnotesize DNN detector with CP = 25\%: a) BER $\times$ SNR curves parameterized on the size of Data Set and Epochs; b) training time period; data-set size is $10^6$.}
\label{i:6}
\end{minipage}
\end{figure}

%...........................
Numerical results in terms of performance {\it vs} complexity trade-off for DNN and ELM are analyzed. The parameters for learning DNN-OFDM and ELM-OFDM architectures are summarized as follows, Fig. \ref{tab:I}.  \\
{\bf OFDM System} -- Transmitter Antenna: $n_T=1$; \, Receiver Antenna:  $n_R= 1$; Modulation Order ($M$-QAM): $M= 4, 16, 32$;  Number of Users:    $U = 4$; \, Sub-carriers:  $N_{c} = 64$; \, 
Pilots Number:   $N_{\rm pilot} = 64, 32, 16$ and $8$;
\, Cyclic Prefix:   $T_g$ = 25\% and 0\%, (Fig. \ref{i:3}); \,
Estimation Methods:   LS e MMSE. \\
\quad{\textbf{Sub-carrier Interference \& Power Allocation}} --
\, Equal Power Allocation (EPA): $P_u(k)=\frac{P_{\textsc{t}}}{N_c}$;
\, Max. \#sub-carriers/user:  $N_{\rm cpu}=$ 16;\, Interfering sub-carriers/user: $N_{\rm cpu}^{\rm eq}$ = 4.\\
\quad{\bf NLoS Channel} -- \, {Cell radius}: 500 m; \, Total Power: $P_{\textsc{t}} = 1$ mW; \, SNR range: $\Bar{\gamma} \in [5;\,\, 25]$ dB;
LoS Channel Model:   Rayleigh; \, Path Loss coefficient:   $\eta = -3$;
\, Path Loss model:   $d_u^{-\eta}$;
\, Coherence Time:   $(\Delta t)_\textsc{c}$ = 5 ms;\\
\quad{\bf DNN Architecture} -- 
\, Hidden Layer:  3;
\,  Input Neuron:   256;
\, Hidden Neurons:   500, 250 and 120;
\, Output Neurons:   64;
Activation Function: 3 Relu and 1 Sigmoid;
Optimizer:   Adam;
Loss Function:   MSE;
\, Epochs:   $10^3$;
\, MCS realizations: $10^4$; \\
\quad{\bf ELM Architecture} -- 
\, Hidden Layer: 1;
\, Input Neurons:   $2$ ;
\, Hidden Neurons:   $L= 50$;
\, Output Neurons:   2;
\, Sub-Network:  64;
Activation Function: Radbas;
\, Pilots symbols:   $I =  50, 100, 200$;
\, Data symbols:   $K=400$;
and \, MCS realizations:  $\mathcal{T}=10^3$.
 \vspace{2mm}

We analyze the influence of several parameters, such as the number of pilots, the number of users, and the SNR training on the BER performance is analyzed. The analytical average BER performance is given by \cite{goldsmith}: $\textrm{BER}^{\textrm{theo}} = \frac{\alpha_\textsc{m}}{2} \begin{bmatrix}  1- \sqrt{\frac{0.5 \beta_\textsc{m} {\Bar{\gamma}}}{1+0.5 \beta_\textsc{m} {\Bar{\gamma}}}} \end{bmatrix},$
where $\alpha_\textsc{m}$ and $\beta_\textsc{m}$ are constants that depend on  the modulation type, {\it i.e.} $\alpha_\textsc{m}$ is the number of nearest neighbors to a constellation at the minimum distance, and $\beta_\textsc{m}$ is a factor relating the minimum distance to the average symbol energy \cite{goldsmith}. For 4-QAM results $\textrm{BER}^{\textrm{theo}} = \frac{1}{2}\left[1- \sqrt{\frac{\gamma_b}{1+ \gamma_b}}\right]$, where $\gamma_b= \frac{{\Bar{\gamma}}}{2}$.

\vspace{1mm}
\noindent{\bf Channel Estimation Task}. In this sub-section, we have analyzed the numerical results related to both channel estimation methods through the normalized mean square error (NMSE) metric,  which can be defined as 
 \begin{equation}\label{eq:NMSE}
{\rm NMSE} = \frac{\sum_{i=1}^{{S}} \left|\mathbf{H}(i) - \tilde{\mathbf{H}}(i)\right|^2}{{S} \, \mathbf{H}_{\textsc{v}} \mathbf{H}_{\textsc{v}}^H}
\end{equation}
where
${\bf H}_{\textsc{v}}={\rm vec}\left(\left[\mathbf{H}(1)\, \mathbf{H}(2) \, \ldots\, \mathbf{H}(S)\right] \right) 
$, with vec($\cdot$) operator indicating the vectorization of a matrix, which converts the $(N_c \times S)$ OFDM channel samples matrix into the ${\bf H}_{\textsc{v}}$ channel frequency domain column vector ($N_c S \times 1$); finally, $S$ is the number of channel estimate samples performed in the channel coherence time interval.  Under medium to high SNR, both linear channel estimators methods are suitable, {\it e.g.}, for  $\Bar{\gamma}=20$ dB, the NMSE$_\textsc{ls} = 52 \times 10^{-4}$, with an order of magnitude in favor or the MMSE method.

\vspace{1mm}
\noindent{\bf Number of Pilots}. The DNN architecture has the advantage of attaining a good performance when compared with the classical channel estimators, such as MMSE and LS, the DNN make-up for the information lack made by interpolation in LS and MMSE methods as depicted in Fig. \ref{i:2}.  The attainable BER for the DNN-OFDM detector is comparable to the BER from MMSE and LS for all SNR regions.  Such a BER is similar to that achieved by the MMSE receiver assuming the same block-type size and arrangement, evidencing that the DNN approach is promising. Also, Fig. \ref{i:2}.(b) reveals that DNN-detector under a reduced number of pilots (8-pilots for DNN {\it vs.} 16-pilots MMSE {\it vs.} 8-pilots LS) outperforms the conventional estimation methods, {\it i.e.}, the DNN-OFDM detector presents desirable robustness against the incomplete pilot's size. 

\begin{table}[!htbp]
\captionof{table}{Simulation Parameters -- Channel, System and DNN and ELM parameters.}\label{tab:I}
\footnotesize
\centering
\begin{minipage}{.48\textwidth}
  \centering
\vspace{-4mm}
\begin{center}
\begin{tabular}{ll}
\toprule
\bf Parameter  & \bf  Value\\
\hline \hline
\multicolumn{2}{c}{\bf OFDM System}\\
\hline
\# Transmitter Antenna  &   $n_T=1$\\
\# Receiver Antenna     &   $n_R= 1$\\
\# Modulation Order ($M$-QAM)    &   $M= 4, 16, 32$\\ 
\# Number of Users      &   $U = 4$\\
\# Sub-carriers         &   $N_{c} = 64$\\
\# Pilots Number        &   $N_{\rm pilot} = 64, 32, 16$ and $8$ \\
\# Cyclic Prefix        &   $T_g$ = 25\% \& 0\%, (Fig. \ref{i:3})\\ 
Estimation Methods      &   LS e MMSE\\ 
\hline
\multicolumn{2}{c}{\textbf{Sub-carrier Interference \& Power Allocation}}\\
\hline
Equal Power Allocation (EPA)                 & $P_u(k)=\frac{P_{\textsc{t}}}{N_c}$\\
Max. \# sub-carriers/user               &  $N_{\rm cpu}=$ 16\\
\# Interfering sub-carriers/user        & $N_{\rm cpu}^{\rm eq}$ = 4\\
\hline
\multicolumn{2}{c}{ \bf NLoS Channel}\\
\hline
\# {Cell radius}        & 500 m\\
Total Power             & $P_{\textsc{t}} = 1$ mW\\
SNR range & $\Bar{\gamma} \in [5;\,\, 25]$ dB\\
Channel Model           &   Rayleigh\\
{Path Loss coef.}       &   $\eta = -3$\\
{Path Loss model}       &   $d_u^{-\eta}$\\
{Coherence Time}        &   $(\Delta t)_\textsc{c}$ = 5 ms\\
\hline
 \label{t:1}
 \end{tabular}
 \end{center}
 \end{minipage}
 \hspace{4mm}
 \begin{minipage}{.48\textwidth}
  \centering
  \vspace{1mm}
 \begin{tabular}{ll}
\hline
\bf Parameter  & \bf  Value\\
\hline 
\multicolumn{2}{c}{\bf DNN Architecture}\\
\hline
\# Hidden Layer         &   3\\
\# Input Neuron         &   256 \\
\# Hidden Neurons       &   500, 250 and 120\\
\# Output Neurons       &   64\\
Activation Function     &   3 Relu and 1 Sigmoid\\
Optimizer               &   Adam\\
Loss Function           &   MSE\\
\# Epochs               &   $10^3$ \\
\# MCS realizations  &   $10^4$\\ 
\hline
\multicolumn{2}{c}{\bf ELM Architecture}\\
\hline
\# Hidden Layer         &   1\\
\# Input Neurons        &   $2$ \\
\# Hidden Neurons       &   $L= 50$\\
\# Output Neurons       &   2\\
\# Sub-Network          &   64\\
Activation Function     &   Radbas\\
\# Pilots symbols       &   $I =  50, 100, 200$\\
\# Data symbols         &   $K=400$\\
\# MCS realizations     &   $\mathcal{T}=10^3$\\ 
\toprule
\end{tabular}
 \end{minipage}
\end{table}

\noindent{\bf Impact of the Number of Users}. So far, we have considered 4 users in the cell; hence, we equally spread $N_{\rm cpu}=N_{c}/U$ sub-carriers per user. However, this model does not challenge the DNN, since, in this scenario, there is no inter-carrier interference. In this way, we have increased the number of active users inside the cell, aiming at verifying how DNN deals with the interference {\it i.e.,} when two users share the same subcarrier. Fig. \ref{fig:difuser} depicts the BER for a cell with $|\mathcal{U}|=5,\dots,8$ users, where eight users mean the maximum co-subchannel interference in a system with 64 sub-carriers where for each user it has been allocated 16 sub-carriers, which results in a maximal number of user per sub-carrier equal to 2.

 \vspace{1mm}
\noindent{\bf Cyclic Prefix Influence on the Estimation Quality}. 
In OFDM systems, the cyclic prefix (CP) is paramount to mitigate inter-symbol interference (ISI); however, it has a cost in terms of power, spectrum, and time. As depicted in Fig. \ref{i:3}, the DNN-based detector holds certain robustness against the absence of CP in low and medium SNR regions due to better BER performance compared to the linear MMSE and LS detectors with no use of CP. This indicates that the analyzed DNN architecture can learn the characteristics of channels and tends to better estimate the channel w.r.t. channel inversion strategy implemented in the LS and MMSE.  The non-use of cyclic prefix is a great advantage offered by ML-based OFDM detectors, resulting in a substantial increment in the overall energy and spectral efficiencies.

\noindent{\bf The SNR$_{\rm train}$ Effect on the BER Performance}. the SNR was set for the training stage, $\textsc{snr}_{\rm train}$ influences notably the BER performance, Fig. \ref{i:4}; {\it e.g.}, at the low SNR data transmission regime, $\Bar{\gamma} = 5$ dB, the DNN-OFDM detector trained under $\textsc{snr}_{\rm train} = 5$ dB attains better performance; similarly, at high SNR regime ($\Bar{\gamma} = 25$ dB), the better performance is reached by the DNN topology trained under $\textsc{snr}_{\rm train} = 25$ dB.

 \noindent{\bf Pilot, Data Set and Modulation sizes}. In the training phase, the DNN deals with the batch size ($\Psi$), and epochs ($\Omega$); such influence on the BER performance is shown in Fig. \ref{i:6}.a), while the training time is depicted in Fig. \ref{i:6}.b).  The better performance is obtained with $\Psi=10000$ and $\Omega=500$, although this demands a high time for training. It is notable that under a low number of batch sizes and epochs the performance is suitable yet, {\it e.g.}, for $\Psi=250$ and $\Omega=1000$, result in a reasonable performance with low training time. 
 
%---------------------------
\noindent{\bf Testing Time}. In contrast with the training time of Fig. \ref{i:6}.b), the {\it testing time}, $i.e.$, the time necessary for each ML-based OFDM detector operate in real scenarios for realizing all detection steps completion.  {The {\it operation time} for both detectors have resulted in $T_\textsc{dnn} = 4.7$ ms and  $T_\textsc{elm} = 4.2$  ms.} \noindent Such an operation time has been measured based on the transmission of a single OFDM frame for DNN, while for ELM, we have considered $I$ = 100 pilots and $K$ = 1 symbol, for all the 64 sub-carriers. As expected, the ELM-OFDM detector presents a lower execution time to {perform} all steps, since its architecture comprises just one single layer.

\noindent{\bf Computational Complexity}  for LS-OFDM, MMSE-OFDM, and ELM-OFDM detectors, in terms of the number of operations parameterized on the number of subcarrier $N_c$,   pilots $I$, and neurons $L$ is given by $\mathcal{O} (I\cdot N_c)$, $ \mathcal{O}(4 \cdot I \cdot N_c^3)$ and $\mathcal{O}(2 \cdot  N_c \cdot L^3)$ respectively. The number of {\it flops} can be computed directly from the implemented code in Matlab using \cite{7}. 

%=====================================
\section{Conclusions}\label{sec:concl}
%======================================
Two ML-based topologies of OFDM detectors have been extensively analyzed and compared with the conventional LS and MMSE detectors, evidencing that such topologies can be more advantageous in terms of performance-complexity trade-off perspective. The DNN-based OFDM detector presented a) a promising BER performance when compared to the classical linear high-complexity inversion matrix-based OFDM detection techniques; b) robustness against the incomplete data training, implying in several pilots reduction; and c) absence of cyclic prefix requirement, increasing the energy and spectral efficiencies of the OFDM system. However, the DNN-based detector architecture results in relatively high computational complexity. The ELM-based OFDM detector has been extensively characterized to overcome this limitation, presenting a superior performance-complexity tradeoff.

\end{document}